\documentclass[graybox]{svmult} 

\usepackage{mathptmx}       
\usepackage{helvet}         
\usepackage{courier}        
\usepackage{type1cm}        
\usepackage{makeidx}         
\usepackage{graphicx}        
\usepackage{multicol}        
\usepackage[bottom]{footmisc}

\usepackage{amsmath, amssymb} 
\usepackage{url}             
\usepackage{enumitem}        

\makeindex             

\begin{document}

\title*{Instilling Organisational Values in Firefighters through Simulation-Based Training}
\author{Nardine Osman\orcidID{0000-0002-2766-3475} and Manel Rodriguez-Soto\orcidID{0000-0003-1339-2018} and Jordi Sabater-Mir\orcidID{0000-0001-6982-3572}} 
\institute{Nardine Osman \at Artificial Intelligence Research Institute (IIIA-CSIC), Barcelona, Spain, \email{nardine@iiia.csic.es} 
\and Manel Rodriguez-Soto \at Artificial Intelligence Research Institute (IIIA-CSIC), Barcelona, Spain, \email{manel.rodriguez@iiia.csic.es} 
\and Jordi Sabater-Mir \at Artificial Intelligence Research Institute (IIIA-CSIC), Barcelona, Spain, \email{jsabater@iiia.csic.es} 
}

\maketitle

\abstract{In firefighting and other emergency operations, decisions made under pressure carry profound ethical weight and can significantly impact incident outcomes and firefighter safety. Traditional training methods, while foundational, often fall short in adequately preparing firefighters for the complex ethical dilemmas and value conflicts inherent in chaotic emergency environments. This paper proposes a conceptual framework for enhancing firefighter training by systematically integrating departmental values into simulation-based training. This approach fosters deeper value internalisation and improves value-driven decision-making under pressure. Furthermore, the underlying tools can also be leveraged to evaluate and refine departmental operational protocols for better alignment with preferred values.} 


\section{Introduction}
\label{sec:1_intro}

In emergency situations, simply following rules and procedures is not enough. Crisis teams often face complex, high-pressure situations where the right course of action isn't always clearly specified by the protocols. In these moments, it's not just technical skills or procedures that matter, but the ability to assess the situation and act in a way that reflects core values. 

Core values are key in many public service professions, like policing, healthcare, education, and social work, where professionals are often required to make difficult decisions that affect people's lives. In firefighting, these values are especially critical. 
They are not just ideals, they are practical tools that guide judgment. 
In critical situations marked by ambiguity and urgency, values serve as a compass, helping firefighters navigate their choices and select those that best protect lives and uphold public trust. 
They serve as practical guides for decision-making in high-pressure situations.

Different fire departments may emphasise different values depending on their culture and community context. 
For example, many fire services in Ecuador traditionally highlight values such as self-sacrifice and abnegation, encouraging firefighters to push beyond their limits in dangerous situations \cite{Bomberos-Quito-Etica}. In contrast, the Catalan Fire Brigade explicitly prioritises firefighter safety as a core value, embedding it firmly into formal ethical guidelines and operational protocols \cite{CAT-Bombers-Values}. 
Even within the same department, individual firefighters can hold different personal values, which can influence how they interpret and respond to situations. 
This variation makes it essential to provide structured training on departmental values to new recruits, as translating these values into consistent actions can be challenging. 

We argue that such structured training is often insufficient or inconsistently implemented today. Despite the acknowledged importance of organisational values, ensuring they are properly put into practice through protocols and followed in personnel's decisions, especially under pressure, remains a significant challenge for many fire services. 
Traditional training often emphasises the transmission of these values through lectures, manuals, and codes of conduct. 
In Catalonia, for example, firefighter trainees engage in classroom sessions where they discuss possible scenarios involving potential value conflicts \cite{CAT-Firefighters-Course}. These discussions aim at helping firefighters reflect on how values apply in practice. 
While essential for establishing foundational knowledge, these methods may not adequately equip firefighters to translate abstract principles into concrete actions during high-stress, ethically challenging situations. The ``knowing-doing gap'', where individuals understand concepts like values but struggle to apply them in real-life situations, is particularly pertinent in emergency response, where the cognitive and emotional pressures of an incident can override rational thought processes if values are not deeply ingrained. 

In the field, firefighters operate under conditions of extreme uncertainty and pressure, which increases the challenge of acting consistently with organisational values. Responders often arrive with incomplete information about the structure, hazards, or victims. Conditions can evolve rapidly and unpredictably, making it difficult to apply predefined rules or values in a straightforward manner. For example, a directive to ``aggressively fight fire'' may be appropriate in one case but dangerously reckless in another where structural collapse is imminent but unconfirmed. Physiological and psychological stress can impair situational awareness, trigger cognitive biases, and lead to emotionally driven decisions. In these moments, ethical dilemmas arise, such as having to choose between firefighter safety and civilian rescue, or deciding whether to follow orders that contradict personal moral judgment. 
These value conflicts demand more than just technical skill. They require the ability to reason ethically, maintain composure under stress, and act in alignment with organisational priorities. Training must therefore address not only cognitive understanding of values but also how values are challenged by stress, emotional strain, and unpredictable operational conditions. 

To overcome this challenge, training must move beyond passive methods and embed values through structured, practice-oriented approaches such as simulation-based training. This paper explores how simulations can support value internalisation and practical decision-making, particularly under pressure.
%
Simulation training offers a promising approach to bridge this knowing-doing gap. It allows firefighters to experience and navigate complex scenarios without real-world risks. 
In these simulations, we introduce specially designed scenarios involving value conflicts. These scenarios allow us to identify which values each firefighter prioritises, as well as provide feedback with guidance on how the firefighter can better align their decisions with their organisation's value preferences. Our analysis of value prioritisation builds on previous work where values were operationalised to enable formal reasoning, allowing us to evaluate how well individual actions and even entire protocols align with those values~\cite{Rodriguez-SotoS25}. 
The result is an AI system that employs a formative process that helps firefighters internalise organisational values 
and strengthens their capacity for value-guided autonomy. 

The rest of this paper is divided as follows: Section~\ref{sec:2_lit_review} provides the background needed to understand this challenge, including a brief literature review on formalising and reasoning with values. Section \ref{sec:4_framework} presents our proposed framework for instilling organisational values in firefighters through simulation-based training. Section \ref{sec:scenario} offers an illustrative scenario to demonstrate the framework's application. 
Finally, Section \ref{sec:8_conclusion} concludes with a summary of key arguments and outlines directions for future research.

\section{Background}
\label{sec:2_lit_review}

Traditional firefighter training has excelled in developing psychomotor skills and procedural knowledge through hands-on training and repetitive drills. These methods are vital for ensuring technical proficiency in tasks like hose deployment, ladder operations, and victim extraction. However, the changing nature of emergency response, with increasingly complex and dynamic incidents such as active shooter situations and wildland-urban interface fires, has highlighted a growing need for enhanced training in non-technical skills. These include critical thinking, dynamic decision-making, communication, teamwork, leadership, and, centrally, ethical reasoning and value-based judgment. While some departments incorporate classroom exercises or scenario-based discussions, there is a growing recognition that more structured methods are needed to effectively develop these higher-order competencies, particularly those related to navigating value dilemmas under pressure.

At the core of this broader training approach are organisational values, which shape the professional culture and identity of the department. For fire services, aligning individual behaviour with core organisational values is essential, not only for effective teamwork, but to act with the integrity and transparency the public expects. Ethical decision-making frameworks such as deontology (duty-based), consequentialism (outcome-based), and virtue ethics (character-based) offer useful perspectives for evaluating decisions under pressure. Most fire departments, in practice, embody a blend of these. Values such as ``integrity'' (a virtue), ``adherence to standard operating procedures'' (a deontological code), and ``prioritising the greatest number of lives saved'' (utilitarianism, a form of consequentialism) frequently appear in mission statements and codes of ethics. 
However, the real challenge lies not in stating these values but in ensuring that they are understood, internalised, and applied, especially under pressure. High-stakes environments are full of value conflicts and pressures that make this alignment difficult. One prominent example is ``noble cause corruption'', where rules are bent or broken in pursuit of what is perceived as a greater good. 

Simulation-based training has emerged as a promising approach in other high-stakes professions such as aviation and medicine. It provides a controlled and replicable environment for trainees to practise skills, make decisions, experience consequences, and receive feedback. 
The evidence base for Simulation-based training's effectiveness in improving technical skills and decision-making in fields like surgery and flight operations is substantial~\cite{nicksa2015innovative,zendejas2013state,socha2016training}. 
We argue that when extended to include the value dimension, it becomes even more powerful: it creates a psychologically safe space in which firefighters can engage with morally complex dilemmas, explore trade-offs, and reflect on their choices. This makes simulation not just a space for technical rehearsal, but a tool for professional and ethical development, which helps bridge the ``knowing–doing'' gap and build long-term value-aligned decision-making capacity. 

While the use of simulation for technical skills training is well-established, its application for value-based education and ethical development remains underexplored, but holds significant promise. To succeed and provide feedback for trainees, such a simulation should be capable of formalising relevant values and reasoning about them. These are challenging tasks that have recently gained traction in artificial intelligence (AI) research. 

Concerning the issue of value representation, values in AI are usually articulated through textual headings or labels (such as `abnegation') without further exploring the concrete relations amongst them, their computational semantics, or importance. In the work of Bench-Capon and Atkinson~\cite{Bench-Capon2009}, an argumentation framework is presented, illustrating how persuasion can rely on the strength of arguments, which in turn depends on social values. The relation between value labels and arguments is assumed to be given and the computational semantics of these values are not addressed. Similarly, Serramia et al.~\cite{SerramiaLR20} automate norm synthesis based on prior knowledge of the connections between specific norms and the values they promote. As with the previous work, the association between norms and value labels is taken as given, and the computational semantics of values is not explored. Initial works have begun exploring the computational semantics of values. Osman and d'Inverno~\cite{OsmanDInverno2024} propose a formal model for value representation, introducing a taxonomy that employs property-based nodes to define the computational semantics of values, while explicitly capturing their interrelationships and relative importance. In some other works, values are treated as special goals that can be formally specified and assessed~\cite{MontesS21,abs-2110-09240}. Reinforcement learning has also been used to learn the computational semantics of values, typically as reward functions or preference models. For example, (Cooperative) Inverse Reinforcement Learning (CIRL) aims to infer a human's reward function (a proxy for their values) from observed behaviour, with CIRL specifically modelling the collaborative nature of this learning process~\cite{10.5555/3157382.3157535}. Similarly, techniques for deep reinforcement learning from human preferences allow AI agents to learn complex behaviours aligned with human values by training on comparisons between behaviour trajectories provided by humans~\cite{10.5555/3294996.3295184}. Rodriguez-Soto et al.~\cite{RodriguezSoto2024VALE} investigates how appropriate mathematical functions can be selected or designed to effectively learn and represent the underlying value semantics. 

In this work, we argue that formally modelling values and defining their computational semantics is essential, as the absence of such semantics critically limits the ability to reason about values. Clear value semantics are necessary for advanced value reasoning and explainability, the foundational elements of effective training. However, due to the lack of existing data on value alignment in the firefighting domain, we rely on the expertise of domain specialists to formalise the relevant values.

As for reasoning about values, we focus here on those approaches that make use of the computational semantics of values. 
In the work of Cranefield et al.~\cite{ijcai2017p26}, reasoning about values is employed to assist agents in making choices about which plans to adopt. Values have states indicating their current level of satisfaction, which evolves over time and influences the agent's plan selection process by reflecting how well the agent's actions or external events align with its targeted priorities. A similar approach is presented by Osman et al.~\cite{OsmanDInverno2024} where the satisfaction of values are defined in terms of targeted priorities (here, referred to as properties) that change with context and over time.  
Instead of linking states to values, Dennis et al.~\cite{DENNIS20161} links actions to values and then uses formal verification to prove that autonomous systems satisfy specific ethical rules. Rules in this work are predefined as ``doing action $a$ in context $c$ counts as a violation of ethical principle $\phi$''. The work of Montes and Sierra~\cite{MontesS21} proposes a value-promoting norm synthesis approach that, in essence, optimises the value alignment mechanism proposed by Sierra et al.~\cite{abs-2110-09240}. In that work~\cite{abs-2110-09240}, value preferences are understood as preferences over world states and the value alignment of a set of given norms is based on the degree to which those norms move us towards preferred states. 

%

A relevant line of work involves the use of reinforcement learning. Multi-objective reinforcement learning (MORL), as demonstrated by Rodriguez-Soto et al.~\cite{Rodriguez-SotoS22}, provides mechanisms for navigating and balancing multiple potentially conflicting values, allowing agents to trade off fairness, utility, and other normative considerations in their policy choices. 
More recently, Rodriguez-Soto et al.~\cite{RodriguezSoto2024AWAI,RodriguezSoto2024VALE,Rodriguez-SotoS25} have applied multi-objective optimization to the medical domain, supporting professionals in evaluating medical actions against bioethical principles (such as beneficence, non-maleficence, autonomy, and justice), and assessing the alignment of medical protocols against these principles.

In this paper, we adopt the MORL framework of Rodriguez-Soto et al.~\cite{Rodriguez-SotoS25} and apply it to firefighter training. We believe this approach can help provide nuanced feedback to trainees by analysing their decisions and highlighting how their actions align with competing values. It can also be used to generate `ideal' pathways that reflect different value-aligned strategies, moving beyond the notion of a single ``right'' response and fostering more reflective and value-aware decision-making. Last, but not least, it can also help fire departments in refining their operational protocols to better promote specific values.

\section{A Framework for Value-Driven Simulation Training for Firefighters}
\label{sec:4_framework}
In this section, we propose a framework for designing, implementing, and leveraging agent-based simulations to instil and reinforce organisational values in firefighters. The framework consists of three key dimensions. First, organisational values must be identified and operationalised. Second, tools must be developed to enable structured reasoning over these values, a necessary foundation for effective value training. Finally, scenarios must be designed and implemented to elicit value-relevant decision-making in realistic, high-stakes contexts. Each of these three dimensions is discussed in further detail in the following subsections, respectively.

\subsection{Identifying and Operationalising Departmental Values}
\label{subsec:4.1_define_ops_values}
Identifying and operationalising departmental values is the foundational phase that builds the ground for value-based reasoning. This phase involves a rigorous process of identifying, understanding, and operationalising the core values that the department wishes to emphasise through simulation training. 

First, there is the issue of value identification and understanding. In the case of firefighting, fire departments usually have their core values clearly identified and articulated in official documents such as mission statements or codes of ethics (e.g. \cite{CAT-Bombers-Values,BCN-Bombers-Values,FDNY-Values}). However, this phase may require more than simply listing values from official documents. It should involve engaging departmental stakeholders in discussions to achieve a shared understanding of what each value means in practical terms within the context of firefighting operations. For example, the value ``professionalism” is often associated with adherence to standard operating procedures but may also encompass maintaining composure, communicating effectively, and engaging in continuous learning.  

Second, there is the critical step of translating these conceptual understandings into clearly defined observable behaviours that can be represented and assessed within a simulated scenario. These observable behaviours formally define how decision-making can be evaluated in terms of upholding or compromising different values. 
For instance, if the value ``teamwork'' is a core value, observable behaviour might include: relaying critical information (such as changes in fire behaviour or victim status) over the radio within a specific time frame, positioning oneself to cover a teammate during a room entry, or responding to a crew member's call for assistance within a defined window. 
Each of these actions can be clearly tracked and assessed within a simulation. Defining values in terms of such actions is what makes values operational. 

As such, the final step is to formalise values in terms of the identified actions. This involves defining clear rules or formulae that specify how particular behaviours promote or undermine specific values. By translating actions into measurable indicators of value alignment, these formal definitions create a structured and objective basis for evaluating decision-making within simulations. They also support consistent and transparent assessment across trainees and scenarios.

This entire phase of identifying and operationalising departmental values naturally presents several challenges. A major difficulty lies in achieving consensus among stakeholders on the operational definitions of values, which is an essential step for ensuring the legitimacy and effectiveness of the resulting training. Reaching such agreement can be difficult due to differing interpretations of abstract values, the inherent complexity of codifying human behaviour, and the need to strike a balance between specificity and the flexibility required to accommodate diverse operational scenarios. Concerning the lack of agreement of different stakeholders, one possible solution is to aggregate stakeholder input systematically following computational social choice theory, as proposed in~\cite{LeraLeriLBJLMRS24}. Another approach to address these challenges is to learn the operationalising of values from observed behaviour or structured stakeholder feedback, as explored in~\cite{RodriguezSoto2024VALE}. However, in the context of firefighting, such behavioural data is often scarce or difficult to collect, making manual elicitation from experts or practitioners the more reasonable route. An illustrative example is presented later in Section~\ref{sec:valuesExample}.

\subsection{Reasoning with Values}
\label{subsec:4.2_value_reasoning}
We now build on the formal value specification introduced above to support structured reasoning over values. 
We propose a formal and computationally grounded framework for evaluating whether an individual's behaviour aligns with multiple organisational values \cite{Gabriel2020}. This challenge, known as the value-alignment problem, refers to ensuring that an agent (human or artificial) acts in ways that reflect the relevant ethical priorities. The problem becomes particularly complex when values conflict or must be respected in dynamic and uncertain environments. To address this, we reformulate the value alignment problem using tools from multi-objective reinforcement learning (MORL), a framework well-suited for decision-making scenarios involving competing objectives \cite{Roijers2017Book,Hayes2021}. 

At the core of our approach is the idea that organisational values, such as ``professionalism'' or ``teamwork'', are formalised as alignment functions that assign a score to each action, indicating how well that action promotes (or demotes) a given value (see subsection~\ref{sec:valuesExample}). The environment in which agents operate is represented as a labelled transition system, which consists of a set of states, a set of actions (or labels), and a set of transitions, where each transition specifies how an action moves the system from one state to another. 
The alignment function of a value evaluates each transition with a score in the range $[-1,+1]$: $1$ represents maximum alignment, $-1$ maximum misalignment, and $0$ indifference. Values between $0$ and $1$ capture partial alignment, while those between $-1$ and $0$ reflect partial misalignment. 

In our proposal, we compute the overall alignment of an agent's behaviour over time by aggregating alignment scores over a sequence of transitions in a multi-objective Markov decision process (MOMDP) \cite{Roijers2017Book}. This expected cumulative alignment corresponds to a value function, a concept from reinforcement learning that quantifies the long-term desirability of possible behaviours \cite{suttonbarto}. 
In the case of our firefighting scenario, the value function is extended to a multi-objective setting: each value (such as professionalism or teamwork) is treated as a separate objective, and the agent receives a distinct reward signal for each one. 
We also identify the \emph{ideal} behaviour or sequences of actions that could have been taken (known as a policy in MOMDP) by applying multi-objective reinforcement learning (MORL) algorithms to compute Pareto-optimal policies (that is, policies that cannot be improved on one value without sacrificing performance on another) \cite{Roijers2017Book}. 
In contexts where a full aggregation of values into a single utility function is infeasible or undesirable, Pareto-optimality offers a principled way to identify the most balanced and defensible behaviours. 

\begin{figure}[!b]
    \centering
    \includegraphics[width=0.5\linewidth]{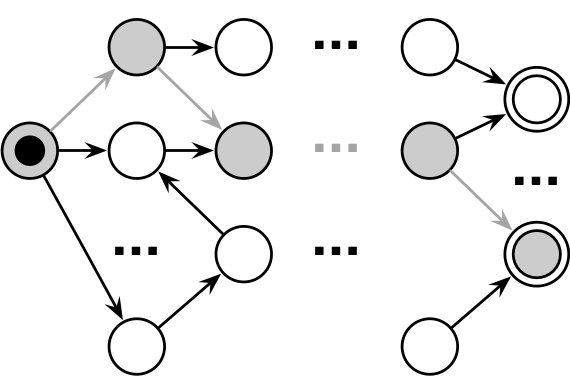}
    \caption{The MOMDP environment is defined through a labelled transitions system, where the initial state is marked with a central black circle, and terminal states are depicted with double circles. 
    At each state, different actions may be available. Each action transitions the system to a new state. Each state encodes relevant features such as fire intensity, occupancy, equipment readiness, visibility, and firefighter health. The grey path illustrates a sample trajectory or policy. It could represent the trainee's chosen sequence of actions or an alternative optimal value-aligned behaviour. It could also provide an illustrative example of where transitions need to be removed or added to improve alignment with organisational values: that is, how permissions and obligations should be added/removed to reshape a departmental protocol in a way that better promotes the intended values and discourages misaligned behaviours.}
    \label{fig:labelled_transition_system}
\end{figure}

Figure~\ref{fig:labelled_transition_system} illustrates how the problem at hand can be modelled as an MOMDP. While our formal value specifications allow us to assess the alignment of individual actions (or transitions in the MOMDP), MORL enable us to assess potential sequences of actions (or policies) to recommend optimal ones. 
Importantly, the same framework also allows us to evaluate paths that were already executed, by measuring how well a chosen sequence of actions aligns with the relevant organisational values. 

By providing feedback on both executed behaviour as well as alternative optimal behaviour, we offer a practical tool for reflection and guided learning that contributes to ingraining organisational values in everyday decision-making. 
Furthermore, by framing the value alignment problem within the formalism of multi-objective reinforcement learning, we provide a rigorous and interpretable approach to ethically grounded decision-making in high-stakes public service professions. 
This approach has been applied in the healthcare domain~\cite{Rodriguez-SotoS25}, and we now extend it to the context of firefighting. 
 
Finally, we extend the proposed MORL-based approach for reasoning about values to focus not only on individual behaviour, but also to evaluate and assess entire departmental protocols. Protocols are standard operating procedures that guide action selection by specifying how firefighters should behave in different states of the world. 
In the context of reinforcement learning, protocols (typically called \emph{policies} \cite{suttonbarto}) indicate which actions are permitted, forbidden, or obligatory in each state. By doing so, they effectively modify the underlying labelled transition system, restricting certain paths (e.g., by introducing new prohibitions) or enabling new ones (e.g., through added permissions or obligations). Evaluating the value-alignment of protocols thus involves analysing the full range of potential behaviours (or paths in the state space) that may emerge when agents follow the protocol, and determining how well these behaviours align with the organisation's values. This is achieved using the same multi-objective reinforcement learning tools introduced earlier, which assess and optimise over the space of possible paths in the labelled transition system.


\subsection{Designing and Implementing Value Training Scenarios}
\label{subsec:4.3_design_scenarios}

After departmental values have been identified and operationalised, and formal tools for reasoning about them have been developed, the next step is designing value training scenarios. These scenarios serve as realistic contexts in which trainees actively engage with and apply organisational values, which reinforces value-based decision-making through simulation-based training.

Effective scenario design should be carried out in close collaboration with stakeholders such as experienced firefighters, training officers, and ethicists to ensure authenticity, relevance, and alignment with learning objectives. Scenarios should reflect realistic firefighting challenges and present genuine ethical dilemmas. These dilemmas can reflect conflicts between organisational values, or even conflicts between personal and organisational values. The richness of these dilemmas is key to fostering meaningful learning. 

To promote deeper cognitive engagement, scenarios must possess sufficient complexity. They should feature branching narratives, allowing the situation to evolve based on the trainee's decisions. This gives trainees a sense of agency and exposes them to the consequences of their choices. To further enrich decision-making under pressure, scenarios can include unexpected events, such as equipment failure or structural collapse, testing trainees' adaptability and value prioritisation under uncertainty. 
Yet, while realism enhances immersion, scenarios must be carefully balanced to avoid excessive cognitive load that could distract from the intended value-learning objectives. 

Insights from computational models that explore value trade-offs \cite{Rodriguez-SotoS25} can inform scenario design by highlighting specific decision points known to create significant value conflicts or by suggesting diverse optimal pathways that represent different value priorities. This can help ensure scenarios are challenging and educationally rich. 

We note that while generative AI tools could significantly assist in generating training scenarios, capturing the nuanced ethical tensions and ensuring practical relevance still requires careful review and input from experienced professionals.  

Once designed, scenarios should be formally implemented as Multi-Objective Markov Decision Processes (MOMDPs). This involves defining the environment (its states, actions, and transitions) as well as introducing the value alignment functions to complete the model. The Pareto Multi-Objective Value Iteration (PMOVI) algorithm \cite{Roijers2017Book} can then be applied to compute the Pareto front, producing alignment scores across all optimal behaviours. The following section provides an illustrative example.

\section{Illustrative Scenario}\label{sec:scenario}
To demonstrate our framework, we present a training scenario involving a simulated firefighting incident. This scenario is based on a real-world inspired environment and has been formalised as a multi-objective Markov decision process (MOMDP) to assess value alignment in firefighter decision-making. It serves to evaluate an agent's alignment with multiple, potentially conflicting organisational values: specifically, the values ``professionalism'' and ``proximity'' of the Catalan Fire Brigade.

In this scenario, a fire breaks out in a multi-storey residential building in a high-density urban neighbourhood. The fire is escalating beyond the initial compartment and there are fears of spreading across multiple units. Several occupants are reported to be inside, with no reliable floor plan available and no accurate information on the number of people trapped. 
Environmental conditions are dynamic: fire intensity evolves stochastically, equipment reliability degrades with use, and partial injuries to firefighters affect decision options. 

The scenario presents a high-stakes situation in which agents must balance strict adherence to procedural safety (Professionalism) with rapid, community-responsive intervention that prioritises human life and public safety (Proximity). The tension between these two values is characteristic of real-world firefighting operations. 

\subsection{Implementing the Scenario as an MOMDP}
Implementing this scenario as a Multi-Objective Markov Decision Process (MOMDP) requires formally specifying both the simulation environment and the value alignment functions.

\subsubsection{Simulation Environment}
We begin by defining the environment as a labelled transition system. This involves specifying a set of states, a set of actions available to the agent (the firefighter), and a transitions that determines how the environment's state evolves in response to those actions. 

Each state represents a snapshot of key environmental variables relevant to the firefighting context, while actions represent the set of possible interventions the firefighter can perform at any given time. The transitions define how each action modifies the current state and determines the resulting next state. 

{\bf States.} 
Each state encodes five key environmental variables: the intensity of the fire in the area to be evacuated, the number of remaining occupants in that area, the firefighter's equipment readiness, their knowledge of the situation (often dictated by visibility), and their current health status. The possible values for each variable are listed below:
\begin{itemize}
    \item Fire intensity: \{None, Low, Moderate, High, Severe\}
    \item Occupancy: \{0, 1, 2, 3, 4\}
    \item Equipment readiness: \{Ready, Not Ready\}
    \item Knowledge (Visibility): \{Good, Poor\}
    \item Health: \{Perfect, Slightly injured, Moderately injured, Incapacitated\}
\end{itemize}

The simulation begins in the following default initial state, although this can be easily modified to suit different training objectives or difficulty levels:
\begin{flalign*}
< \; & Fire intensity = Moderate, \\
& Occupancy = 4,\\
& Equipment = Not Ready,\\
& Knowledge = Poor, \\
& Health = Perfect \; >
\end{flalign*}

We currently model the simulation using a single firefighter agent to maintain clarity and computational tractability in this initial implementation. The framework, however, is designed to accommodate future extensions involving multiple interacting agents. In the current setting, the firefighter makes a sequence of decisions (actions), each of which alters one or more variables of the environment's state.

The simulation terminates under one of two conditions: either all occupants are safely evacuated and the fire is fully extinguished (a successful outcome), or the firefighter becomes incapacitated (an unsuccessful outcome).

To maintain conceptual clarity and computational feasibility at this stage, we intentionally keep the state variables and their domains relatively simple. Nonetheless, the resulting state space is already non-trivial, comprising a total of $5 \cdot 5 \cdot 2 \cdot 2 \cdot 4 = 400$ unique states. Some of these are terminal states, meaning they have no further transitions and represent final outcomes in the scenario. 

{\bf Actions and Transitions.} 
At each state, the firefighter trainee can choose one of the following five available actions. These actions update the state of the simulation to reflect the consequences of decisions made in a changing environment. 
\begin{itemize}
    \item \textbf{Evacuate Occupants}: Attempts to remove one occupant from the affected area. If performed with poor knowledge of the situation (\textit{Knowledge=Poor}) or when the equipment is not ready (\textit{Equipment=NotReady}), the firefighter's \textit{Health} status decreases by one level (e.g., from Slightly Injured to Moderately Injured). Additionally, if this action is taken when the fire intensity is Moderate or higher, health is also decreased by one level. Under severe fire intensity, the equipment becomes degraded (i.e., \textit{Equipment=NotReady}).
    \item \textbf{Contain Fire}: Reduces the fire intensity by one level. For instance, a \textit{Moderate} fire becomes \textit{Low}. This represents a safer action for controlling fire progression.
    \item \textbf{Aggressive Fire Suppression}: Reduces the fire intensity by two levels. For instance, a \textit{Moderate} fire becomes \textit{None}. 
    If performed with poor knowledge of the situation or when the equipment is not ready, the firefighter's \textit{Health} status decreases by one level. Additionally, if this action is taken when the fire intensity is Moderate or higher, health is also decreased by one level. Under severe fire intensity, the equipment becomes degraded. 
    \item \textbf{Prepare Equipment}: Sets \textit{Equipment Readiness} to \textit{Ready}, allowing safer execution of subsequent tasks.
    \item \textbf{Update Knowledge}: Sets \textit{Knowledge} to \textit{Good}, reflecting improved situational awareness (e.g., clearing smoke, updated floor plan, or obtaining new intelligence).
\end{itemize}

Each action induces a transition from the current state to a new state, governed by rules depending on the context (as illustrated above). These transitions form the backbone of the simulation's decision-making dynamics and are key to evaluating value alignment.

\subsubsection{Value Specification}\label{sec:valuesExample}
In consultation with members of the Catalan Fire Brigade, we identified two key organisational values that frequently come into tension during real-world operations and are particularly relevant to this scenario: \textit{professionalism} and \textit{proximity}.

Professionalism is defined as adherence to the organisation's formal rules and procedures. In the context of the Catalan Fire Brigade, this includes prioritising firefighter safety over aggressive or high-risk tactics. Actions that reduce operational risk, such as preparing equipment before entering hazardous zones or maintaining clear situational awareness, are considered strongly aligned with this value.

Proximity, by contrast, refers to the firefighter's understanding of the societal and territorial context in which the incident occurs. It reflects a sensitivity to the impact of the event on the local community and an emphasis on rapid, responsive intervention when lives or livelihoods are at stake. Proximity may justify a more urgent and risk-tolerant approach, such as attempting rescue operations under suboptimal conditions, when delaying could result in fatal consequences or significant social disruption. 

Tensions between these two values emerge when the requirements of procedural safety conflict with the immediate needs of the community. For example, entering a burning building without adequate preparation may violate protocols (compromising Professionalism) but could align with Proximity if it is the only chance to save trapped occupants.

To formally capture these dynamics, we associate each value with an \textit{alignment function}: a mapping from transitions (that is, state-action pairs) to a numerical alignment score in the range $[-1,+1]$. A score of $+1$ indicates strong alignment with the value, $0$ denotes irrelevance or neutrality, and $-1$ represents direct misalignment. Importantly, alignment functions in this scenario do not evaluate the outcomes of actions, but rather assess whether the actions themselves align with the organisation's expectations (such as following safety protocols or responding appropriately to community needs) given the specific context in which they are taken.

These alignment functions enable values to be operationalised and quantitatively assessed. 
They allow us to evaluate how well individual decisions reflect organisational priorities. 

In what follows, we define how each action in our scenario promotes or demotes these values, depending on the state in which it is performed.

\begin{itemize}
    \item \textbf{Evacuate Occupants:}\\
    \textit{Professionalism:} This action is strongly aligned with professionalism ($+1$) when fire intensity is \textit{None} and situational knowledge is \textit{Good}, indicating that the firefighter is acting within safe conditions. As fire intensity increases and knowledge of the situation (such as visibility) decreases, alignment gradually decreases, reaching neutrality ($0$) under conditions of \textit{Severe} fire and \textit{Poor} knowledge. If this action is taken when no occupants are present, it results in maximum misalignment ($-1$), reflecting a critical error in judgment leading to a waste of valuable time.\\
    \textit{Proximity:} Evacuating occupants is strongly aligned with proximity ($+1$) whenever occupants are present. However, performing this action when no occupants remain results in maximum misalignment ($-1$), as it wastes critical time and resources.

    \item \textbf{Contain Fire:}\\
    \textit{Professionalism:} Containing fire is strongly aligned with professionalism ($+0.8$), reflecting a controlled and measured response. Attempting this action when fire intensity is \textit{None} results in maximum misalignment ($-1$), as it indicates an unnecessary action.\\
    \textit{Proximity:} This action is weakly aligned with proximity ($+0.2$), since it contributes to overall safety but does not directly prioritise occupant rescue. As with professionalism, applying it when fire intensity is \textit{None} results in maximum misalignment ($-1$).
    
    \item \textbf{Aggressive Fire Suppression:}\\
    \textit{Professionalism:} This action is moderately aligned with professionalism ($+0.6$) under normal conditions, but alignment drops (to $+0.3$) if performed without equipment readiness, reflecting increased operational risk. Executing this action when fire intensity is \textit{None} yields maximum misalignment ($-1$).\\
    \textit{Proximity:} Aggressive fire suppression is moderately aligned with proximity ($+0.5$), as it prioritises rapid control of the fire. However, if executed when fire intensity is \textit{None}, it results in maximum misalignment ($-1$), as it fails to support actual community needs.
    
    \item \textbf{Prepare Equipment:}\\
    \textit{Professionalism:} Preparing equipment is moderately aligned with professionalism ($+0.5$). However, if the equipment is already ready, performing this action leads to maximum misalignment ($-1$), signalling wasted time and effort.\\
    \textit{Proximity:} From the perspective of proximity, this action is slightly misaligned ($-0.1$), as it delays urgent rescue or fire suppression activities. When equipment is already ready, performing this action yields maximum misalignment ($-1$). 
    
    \item \textbf{Update Knowledge:}\\
    \textit{Professionalism:} Updating knowledge is strongly aligned with professionalism ($+1$), as it supports informed decisions that follow protocol. If performed when knowledge is already \textit{Good}, it is considered a waste of time and results in maximum misalignment ($-1$).\\
    \textit{Proximity:} This action is moderately misaligned with proximity ($-0.5$), as it delays direct intervention to rescue occupants. If knowledge is already \textit{Good}, performing this action again leads to maximum misalignment ($-1$).
\end{itemize}

\subsection{Scenario Execution and Feedback}
Once the scenario is fully modelled as a multi-objective Markov decision process (MOMDP), execution involves having the trainee interact with the simulation environment through a sequence of state-action transitions. Each simulation episode begins from a defined initial state and proceeds until a terminal state is reached: either the successful evacuation of all occupants and suppression of the fire (desired outcome), or the incapacitation of the firefighter (undesired outcome). 
At each decision point, the agent (trainee) selects an action based on their perception of the current state. The environment responds by transitioning to a new state (sometimes in a stochastic manner) and updating relevant variables such as fire intensity, equipment readiness, and health. Simultaneously, alignment scores for each organisational value (in this case, professionalism and proximity) are computed using the predefined alignment functions over the state-action pair.

The cumulative alignment score generated over the course of the simulation offers a quantitative assessment of how well the trainee's behaviour aligns with the fire department's core values. This provides a \textbf{behaviour assessment} that helps trainees (and training officers) understand which values the trainee is prioritising in practice and how well those priorities align with the department's stated priorities.

In addition to this score, the system computes one or more pareto-optimal policies (or paths that the trainee could have taken) using pareto multi-objective value iteration (PMOVI). This provides \textbf{feedback on ideal behavioural trajectories} that prioritise values in a way that aligns with the department's preferences. 

In summary, the simulation-based training helps trainees explore complex value conflicts and experience the consequences of their decision making. Behaviour assessment, coupled with feedback on ideal behaviour, helps bridge the knowing–doing gap and and promote value-aligned decision making under pressure in real-world firefighting scenarios.

\section{Conclusion}
\label{sec:8_conclusion}
The proposed framework addresses a critical gap in firefighter training by aiming at systematically ingraining values in firefighter's decision making. Through simulations, firefighters are exposed to realistic and complex scenarios with value conflicts, promoting deeper reflection on values. Over time, repeated engagement fosters deeper understanding and intuitive application of organisational values. 

%

Our current computational model already supports nuanced feedback on value alignment by assigning concrete alignment scores to actions and trajectories (e.g., scoring a path as 0.8 for Professionalism and 0.4 for Proximity). This enables trainees to receive targeted insights into how their decisions aligned or conflicted with multiple values. For example, the system might explain: “Your decision to enter immediately showed strong commitment to victim rescue (Proximity), but doing so before confirming team readiness presented a challenge to firefighter safety (Professionalism).” This kind of detailed value-aware feedback promotes structured discussion about trade-offs, helping deepen learning and better prepare trainees for ethically complex real-world decisions.

While the framework has been fully implemented and firefighter actions have been simulated, it has not yet been tested in real training settings. We look forward to testing our proposed training framework with the Catalan Fire Brigade to evaluate its impact and gather feedback from both trainees and training officers. 

As a direction for future work, the use of large language models (LLMs) could further enhance the feedback process by generating human-readable explanations of value conflicts and highlighting how trainee decisions reflect different value priorities. 

Finally, the potential impact of this approach extends beyond firefighting. Many high-stakes professions such as law enforcement, healthcare, and emergency management face similar challenges of instilling core values and supporting ethical decision-making under pressure. The principles of operationalising values and designing dilemma-rich scenarios for simulation-based training offer a pragmatic and effective complement to traditional case-based or lecture-driven ethics instruction.


\bibliographystyle{plain}
\bibliography{references}

\end{document}